\input{aipcheck}

\documentclass[
    ,final            
    ,numberedheadings 
  ]
  {aipproc}

\usepackage{array}
\layoutstyle{6x9}

\def\gad{\hat{\gamma}}

\def\lesssim{\mathrel{\rlap{\lower 4pt \hbox{\hskip 1pt $\sim$}}\raise 1pt\hbox {$<$}}}
\def\gtrsim{\mathrel{\rlap{\lower 4pt \hbox{\hskip 1pt $\sim$}}\raise 1pt\hbox {$>$}}}

\newcommand{\dsfrac}[2]{\displaystyle{\frac{#1}{#2}}}

\newcommand{\secref}[1]{Section~\ref{#1}}
\newcommand{\figref}[1]{Fig.~\ref{#1}}

\begin{document}

\title{Dynamical efficiency of collisionless magnetized shocks in
  relativistic jets}


\author{Miguel A. Aloy}{
  address={Departamento de Astronom\'ia y Astrof\'isica, Universidad
    de Valencia, 46100, Burjassot, Spain}, email=miguel.a.aloy@uv.es
}

\author{Petar Mimica}{}

\begin{abstract}
  The so-called internal shock model aims to explain the light-curves
  and spectra produced by non-thermal processes originated in the flow
  of blazars and gamma-ray bursts.  A long standing question is
  whether the tenuous collisionless shocks, driven inside a
  relativistic flow, are efficient enough to explain the amount of
  energy observed as compared with the expected kinetic power of the
  outflow. In this work we study the dynamic efficiency of conversion
  of kinetic-to-thermal/magnetic energy of internal shocks in
  relativistic magnetized outflows. We find that the collision between
  shells with a non-zero relative velocity can yield either two
  oppositely moving shocks (in the frame where the contact surface is
  at rest), or a reverse shock and a forward rarefaction. For
  moderately magnetized shocks (magnetization $\sigma\simeq 0.1$), the
  dynamic efficiency in a single two-shell interaction can be as large
  as $40\%$. Hence, the dynamic efficiency of moderately magnetized
  shocks is larger than in the corresponding unmagnetized two-shell
  interaction. We find that the efficiency is only weakly dependent on
  the Lorentz factor of the shells and, thus internal shocks in the
  magnetized flow of blazars and gamma-ray bursts are approximately
  equally efficient.
\end{abstract}

\maketitle


\section{Introduction}

Internal shocks (ISs) \citep{Rees:1994ca} are invoked to explain the
variability of blazars \citep[e.g.,][]{Spada:2001p815,Mimica:2005sp}
and the light curves of the prompt phase of gamma-ray bursts (GRBs)
\citep{Sari:1995oq,Sari:1997p766,Daigne:1998wq}. A long standing
concern in this model is the question whether it is efficient enough
to explain the relation between the observed energies both in the
prompt GRB phase and in the afterglow \citep[see,
e.g.,][]{Kobayashi:1997p657,Beloborodov:2000p632,Kobayashi:2000p599,Fan:2006p1375}. To
assess the efficiency of the internal shock model, most of the previous
works focus on the comparison between the observed light curves and
the model predictions employing a simple inelastic collision of
two-point masses
\cite{Kobayashi:1997p657,Lazzati:1999p1360,Nakar:2002p1323,Tanihata:2003p1291,Zhang:2004p1381}. Less
attention has been paid to the hydrodynamic effects during the shell
collision \citep[but
see,][]{Kobayashi:2000p599,Kino:2004p811,Mimica:2004fy,Mimica:2005sp,Bosnjak:2009p1427}.

The ejecta in GRBs and blazars may be significantly magnetized,
particularly if they originate from a Poynting-flux-dominated flow
\cite{Usov:1992hp}. Forming shocks in highly magnetized media is
challenging since the Alfv\'en speed approaches the speed of
light. Therefore, to account for the observed phenomenology it is
necessary to address how efficient can be shock dissipation of the
internal collisions in arbitrarily magnetized flows. This question has
been partly considered by a few recent works
\citep[e.g.,][]{Fan:2004p1007,Mimica:2007db}, and only very recently
has been studied extensively \cite{Mimica:2010Al}.

The base for a study the efficiency of internal collisions is the
determination of the dynamic efficiency of a single binary collision,
i.e., the efficiency of converting the kinetic energy of the colliding
fluid into thermal and/or magnetic energy. Thus, we model ISs as
results of the collision of (magneto-)hydrodynamic shells of plasma
with a non-zero relative velocity. The contact surface, where the
interaction between the shells occurs, can break up either into two
oppositely moving shocks (in the frame where the contact surface is at
rest), or into a reverse shock and a forward rarefaction. The
determination of whether one or the other possibility occurs is
computed by estimating the invariant relative velocity between the
fastest and the slowest shell, i.e., by solving the relativistic
magnetohydrodynamic (RMHD) Riemann problem posed by the piecewise
uniform states given by the physical quantities on the two interacting
shells (Sect.~\ref{sec:riemann}). In \secref{sec:edissipation} we
define precisely the notion of dynamic efficiency, both for shocks and
rarefactions.  We perform a parametric study of the binary shell
collision dynamic efficiency in \secref{sec:parametric}. Finally, the
discussion and conclusions are listed in \secref{sec:discussion}.

\section{RMHD Riemann problem}
\label{sec:riemann}

The interaction between parts of the outflow with varying properties
can be modeled in terms of Riemann problems, i.e. relativistic
magnetohydrodynamic initial-value problems with two constant states
separated by a discontinuity. This approach allows us to use our
models to sample very finely a large parameter space. The approach is
also justified by the fact that the flow is cold and
ultrarelativistic, so that its lateral expansion is negligible. Thus,
a description of the interactions assuming planar symmetry suffices to
compute the dynamic efficiency of such interactions.

In the following we use subscripts $L$ and $R$ to denote properties of
the (faster) left and (slower) right state, respectively. We normalize
the rest-mass density $\rho$ to $\rho_R$, the energy density to
$\rho_R c^2$ ($c$ is the speed of light) and the magnetic field
strength to $c\sqrt{4\pi \rho_R}$.

For the initial thermal pressure of both states we assume that it is a
small fraction of the density, $p_L = \chi \rho_L$ and $p_R = \chi$,
and that magnetic fields are perpendicular to the direction of the
flow propagation.  The remaining parameters determining the RMHD
Riemann problem are: the density contrast $\rho_L$, the Lorentz factor
of the right state $\Gamma_R$, the relative Lorentz factor difference
$\Delta g := (\Gamma_L - \Gamma_R)/\Gamma_R$, and the magnetizations
of left and right states, $\sigma_L := B_L^2/(\Gamma_R^2(1+\Delta
g)^2\rho_L)$ and $\sigma_R := B_R^2/\Gamma_R^2$, where $B_L$ and $B_R$
are the lab frame magnetic field strengths of left and right states,
respectively. Furthermore, we define the total (thermal + magnetic)
pressure, the total specific enthalpy and the total energy density,
respectively, as
\begin{eqnarray}
p^* := p + \dsfrac{B^2}{2\Gamma^2} = p + \dsfrac{\sigma\rho}{2}\, ,\\
h^*:=  1 + \epsilon + p / \rho + \sigma \, , \label{eq:hstar} \\
e^*:= \rho (1 + \epsilon) + \dsfrac{\sigma\rho}{2}\, ,\label{eq:estar}
\end{eqnarray}
where $\epsilon$ denotes the specific internal energy.

The typical structure of the flow after the break-up of the initial
discontinuity consists of the two initial states, and two intermediate
states separated by a contact discontinuity (CD) -see
\cite{Romero:2005zr}. The total pressure and velocity are the same on
both sides of the CD. The quantity $\sigma/\rho$ is uniform
everywhere, except across the CD, where it can have a jump.  We denote
the total pressure of intermediate states $p_S^*$, and rest-mass
density left and right of the CD as $\rho_{S,L}$ and $\rho_{S,
  R}$\footnote{In the context of ISs, if the flow is ultrarelativistic
  in the direction of propagation, the velocity components
  perpendicular to the flow propagation should be negligibly small
  and, hence, they are set up to zero in our model. If such velocities
  were significant, appreciable changes in the Riemann structure may
  result as pointed out in \citet{Aloy:2006p2560} or
  \citet{Aloy:2008kx}.}.
 
One of the key steps in solving a Riemann problem is to determine
under which conditions shocks can form. This happens when the
Lorentz-invariant relative velocity between the left and right states,
measured in the frame of the CD, $v_{lr} := (v_l - v_r)/(1 - v_lv_r)$
is larger than the limiting value \cite{Mimica:2010Al}
\begin{equation}\label{eq:V_LR2S}
(v_{lr})_{2S} = \left\{
  \begin{array}{rl}
    \sqrt{\dsfrac{(p_L^* - p_R^*)(e_{S,R}^*(p_L^*) - e_R^*)}{(e_{S,R}^*(p_L*) + p_R^*)(e_R^* + p_L^*)}} & \mathrm{if}\ p_L^*=p_S^*>p_R^*\\[4mm]
    \sqrt{\dsfrac{(p_R^* - p_L^*)(e_{S,L}^*(p_R^*) - e_L^*)}{(e_{S,L}^*(p_R^*) + p_L^*)(e_L^* + p_R^*)}} & \mathrm{if}\ p_L^*<p_R^*=p_S^*
  \end{array}
  \right.
\end{equation}

Generally, the quantity $(v_{lr})_{2S}$ can be only determined
numerically.  If $(v_{lr}) < (v_{lr})_{2S}$, a single shock and a
rarefaction emerge from the initial discontinuity.

\section{Energy dissipation efficiency of ISs}
\label{sec:edissipation}

To study ISs we idealize interactions of parts of the outflow moving
with different velocities as collisions of homogeneous shells.  In our
model the faster (left) shell catches up with the slower (right) one
yielding, in some cases, a pair of shocks propagating in opposite
directions (as seen from the CD frame).  In order to cover a wide
range of possible flow Lorentz factors and shell magnetizations, we
assume that initially, the flux of energy in the lab frame is uniform
and the same in both shells (see \cite{Mimica:2010Al}). We then
compute the break up of the initial discontinuity between both shells
using the exact Riemann solver developed by \citet{Romero:2005zr},
assuming an ideal gas equation of state (EoS) with an adiabatic index
$\gad = 4/3$.

\paragraph{Efficiency of energy dissipation by a shock}
To model the dynamic efficiency of energy dissipation we follow the
approach described in \cite{Mimica:2010Al}. By using the exact solver
we determine the existence of shocks and (in case one or two shocks
exist) obtain the hydrodynamic state of the shocked fluid. We use
subscripts $S, L$ and $S, R$ to denote shocked portions of left and
right shells, respectively. In the following we treat the efficiency
of each shock separately.

{\bf Reverse shock. } To compute the efficiency we need to compare the
energy content of the initial (unshocked) faster shell with that of
the shocked shell at the moment when RS has crossed the initial
shell. Assuming an initial shell width $\Delta x$, we define total
initial kinetic, thermal and magnetic energy \cite{Mimica:2010Al}
\begin{equation}\label{eq:energies}
  \begin{array}{rcl}
    E_K(\Gamma, \rho, \Delta x) &:=&  \Gamma (\Gamma - 1) \rho \Delta
    x\, ,\\
    E_T (\Gamma, \rho, p, \Delta x) &:=&  [(\rho \varepsilon + p)
    \Gamma^2 - p] \Delta x\, ,\\
    E_M (\Gamma, \rho, \sigma, \Delta x) &:=& \left(\Gamma^2 - 1/2\right)\rho \sigma \Delta x\, .
 \end{array}
\end{equation}
When the RS crosses the whole initial shell, the length of the
compressed shell (i.e., the fluid between the RS and the CD) is
$\zeta_L \Delta x$, where $\zeta_L: =(v_{CD} - v_{S, L})/(v_L - v_{S,
  L}) < 1$, and $v_{CD}$ and $v_{S, L}$ are velocities (in the lab
frame) of the contact discontinuity and the RS, both obtained from the
solver.  We normalize the energies taking $\Delta x=1$, and define
the dynamic \emph{thermal} and \emph{magnetic} efficiencies, i.e., the
fraction of the initial energy that the RS has converted into thermal
and magnetic energy, respectively, as
\begin{eqnarray}
  \varepsilon_{T, L} := \dsfrac{E_T(\Gamma_{S, L}, \rho_{S, L}, p_{S, L}, \zeta_L) -
    E_T(\Gamma_R (1 + \Delta g), \rho_L, \chi \rho_L, 1)}{E_0}\, ,\label{eq:thermalL}\\
  \varepsilon_{M, L}:= \dsfrac{E_M(\Gamma_{S, L}, \rho_{S, L},
    \sigma_{S, L}, \zeta_L) - E_M(\Gamma_R (1 + \Delta g), \rho_L,
    \sigma_L, 1)}{E_0}\, ,  \label{eq:magneticL}
\end{eqnarray}
where $E_0$ is the total initial energy of both shells 
\begin{equation}\label{eq:E_0}
  \begin{array}{rl}
    E_0 :=& E_K(\Gamma_R  (1 + \Delta g), \rho_L, 1) +
    E_K(\Gamma_R, 1, 1) + E_T(\Gamma_R (1 + \Delta g), \rho_L, \chi \rho_L, 1)+\\[4mm]
    & E_T(\Gamma_R, 1, \chi, 1) +  E_M(\Gamma_R (1 + \Delta g), \rho_L,
\sigma_L, 1) +E_M(\Gamma_R, 1, \sigma_R, 1)\, .
    \end{array}\, 
\end{equation}

{\bf Forward shock. } In complete analogy, we define the thermal and
magnetic efficiencies for the forward shock,
\begin{equation}\label{eq:thermalR}
  \varepsilon_{T, R} := \dsfrac{E_T(\Gamma_{S, R}, \rho_{S, R}, p_{S, R}, \zeta_R) - E_T(\Gamma_R, 1, \chi, 1)}{E_0}\, ,
\end{equation}
\begin{equation}\label{eq:magneticR}
  \varepsilon_{M, R}:= \dsfrac{E_M(\Gamma_{S, R}, \rho_{S, R}, \sigma_{S,
      R}, \zeta_R) - E_M(\Gamma_R, 1, \sigma_R, 1)}{E_0}\, ,
\end{equation}
where $\zeta_R := (v_{S, R} - v_{CD})/(v_{S, R} - v_R) < 1$, and
$v_{S, R}$ is the lab frame velocity of the FS.


Combining equations \eqref{eq:thermalL}, \eqref{eq:magneticL},
\eqref{eq:thermalR} and \eqref{eq:magneticR} we define the dynamic
thermal and magnetic efficiency of ISs
\begin{equation}\label{eq:thermal-magnetic}
  \varepsilon_T := \varepsilon_{T, L} + \varepsilon_{T, R}\:,\:\:\:\:\:
  \varepsilon_M:= \varepsilon_{M, L} + \varepsilon_{M, R}\, .
\end{equation}

\paragraph{Efficiency of energy dissipation by a rarefaction}
In a rarefaction there is a net conversion of magnetic and/or thermal
energy into kinetic energy, thus the net dynamic efficiency produced
by a rarefaction, defined as in e.g., Eq.~\eqref{eq:magneticL}, should
be negative. Therefore, it is possible that in some cases the total
(left plus right) thermal or magnetic efficiency
(Eqs.~\eqref{eq:thermal-magnetic}) becomes negative. However, this
situation does not correctly model the fact that, in cases where a
single shock exists, it is still able to radiate away part of the
thermal or magnetic energy behind it. Thus, we set $\varepsilon_{T, L}
= \varepsilon_{M, L}=0$ ($\varepsilon_{T, R} = \varepsilon_{M, R}=0$)
if the reverse (forward) shock is absent.

\section{Parametric study of the dynamic efficiency}
\label{sec:parametric}

\begin{figure}
\begin{tabular}{m{.5\linewidth}m{.5\linewidth}}
(a) & (b) \\*\vspace{-0.9cm}
\includegraphics[width=7.5cm]{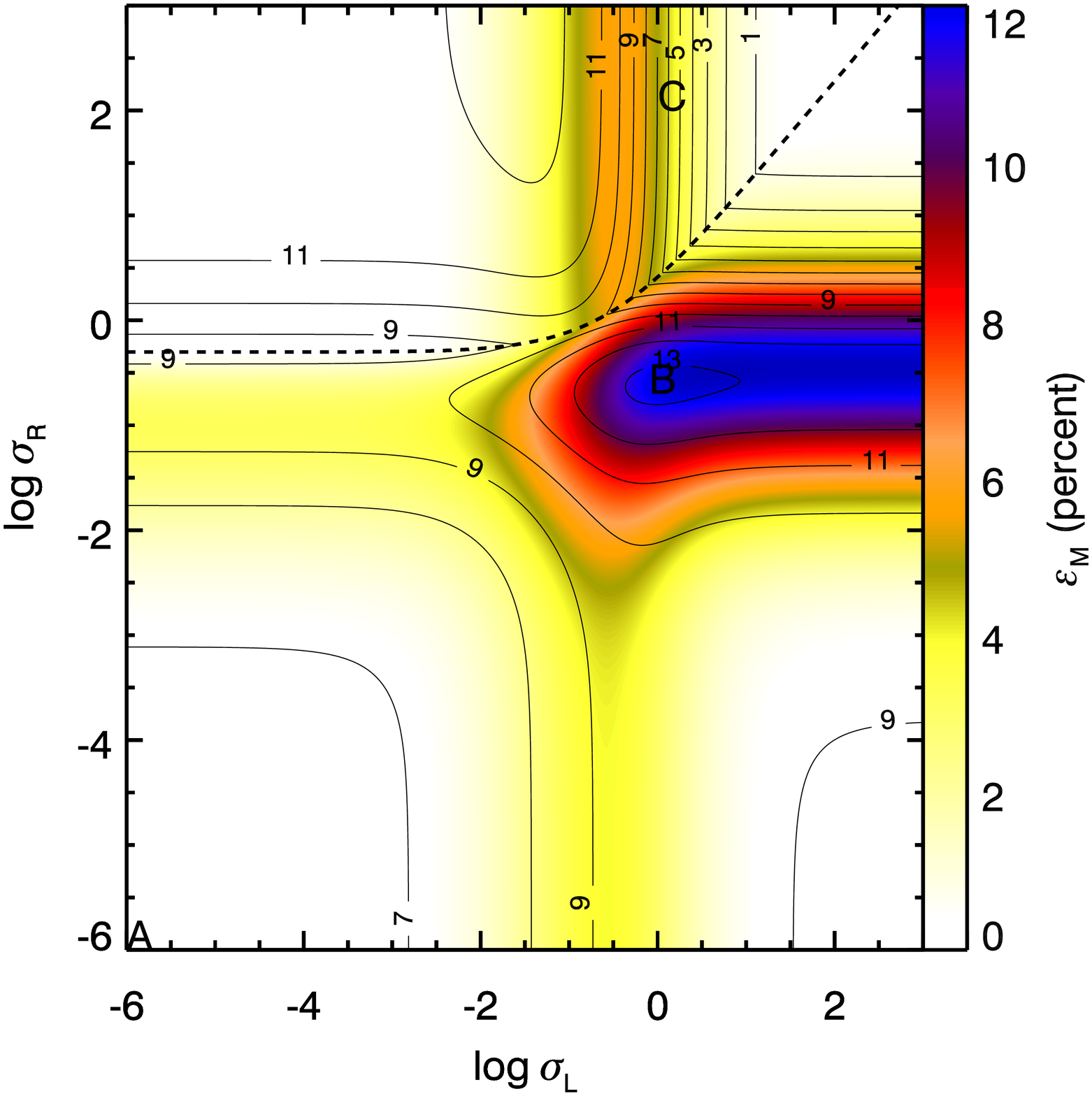} &\vspace{-0.9cm}
\includegraphics[width=7.5cm]{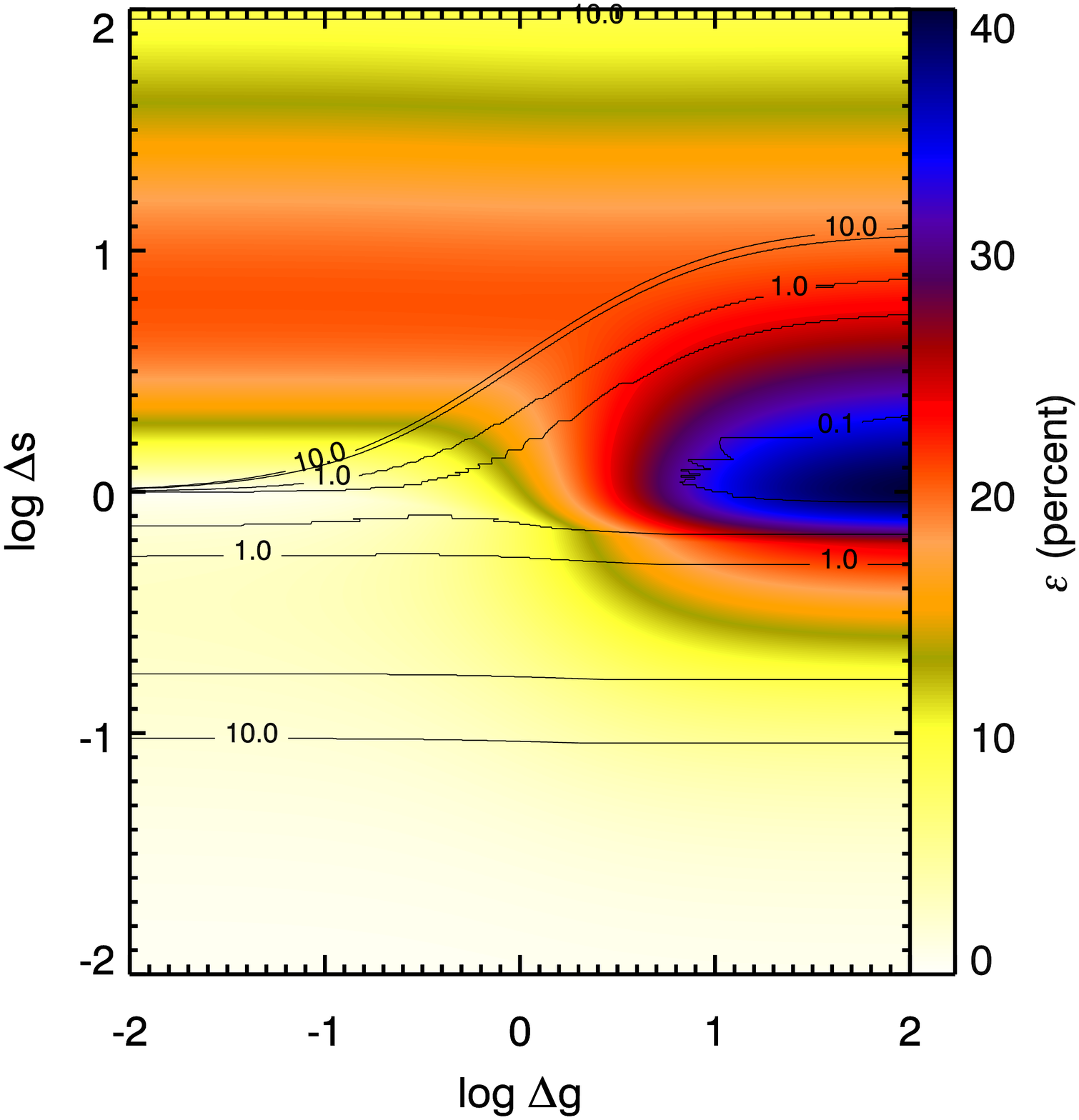}
\end{tabular}
\caption{{\em Panel (a)}: contours of the total dynamic efficiency
  $\varepsilon_T + \varepsilon_M$ (Eq.~\eqref{eq:thermal-magnetic}) in
  the GRB regime ($\Gamma_R = 100$, $\Delta g = 1$; left panel) for
  different combinations of $(\sigma_L, \sigma_R)$. Contours indicate
  the efficiency in percent and their levels are $1$, $2$, $3$, $4$,
  $5$, $6$, $7$, $8$, $9$, $10$, $11$, $12$ and $13$. In the region of
  the parameter space above the dashed line there is no forward shock,
  while the reverse shock is always present for the considered
  parametrization. Colors: magnetic efficiency $\varepsilon_M$ in
  percent. {\em Panel (b)}: The color scale indicates the value of the
  maximum total dynamic efficiency (in percent) as a function of the
  parameter pair $(\Delta g, \Delta s)$. The values of the rest of the
  parameters are fixed to $\Gamma_R=100$, and $\chi=10^{-4}$.
  Contours: magnetization of the slowest shell: $\sigma_R=0.1$, $0.5$,
  1, 5 and 10.}
\label{fig:blazar}
\end{figure}

Next we study the dynamic dissipation efficiency in the process of
collision of cold, magnetized shells. The shells are assumed to be
cold because in the standard fireball model, almost all the internal
energy of the ejecta has been converted to kinetic energy {\it before}
ISs happen. If the ejecta were accelerated by magnetic fields, then
the flow is cold (i.e., $\chi \ll 1$) all the way from the beginning
to the internal shock phase. Thus, we fix $\chi = 10^{-4} \ll 1$, to
model initially cold shells. We set $\Delta g = 1$ as a reference
value, and consider the dependence of the total dynamical efficiency
($\varepsilon_T$ + $\varepsilon_M$) on the individual magnetization of
each shell $\sigma_L$ and $\sigma_R$ \figref{fig:blazar}(a). For the
purpose of this study we take $\Gamma_R=100$, which is a typical value
of GRB outflows. Smaller values, $\Gamma_R=10$, more representative of
blazar outflows, do not yield significant differences in the total
dynamical efficiency \cite{Mimica:2010Al}.

The maximum efficiency (where $\varepsilon_T+\varepsilon_M \gtrsim
0.13$) is attained for ($\sigma_R, \sigma_L) \approx (0.2, 1)$. In the
region above the dashed line of \figref{fig:blazar}(a) the FS is
absent and, thus, since only the RS dissipates the initial energy, the
efficiency slightly drops. However, the transition between the regime
where the two shocks operate or only the RS exists is smooth. The
reason being that the efficiency below the separatrix of the two
regimes and close to it is dominated by the contribution of RS. As
expected, when either $\sigma_L$ or $\sigma_R$ approach low values,
the dynamic efficiency ceases to depend on them.

We illustrate the differences in terms of dynamical efficiency by
choosing three characteristic models in the parameter space (letters
$A$, $B$ and $C$; \figref{fig:blazar}a). Model $A$ corresponds to a
prototype of interaction between non-magnetized shells $(\sigma_L =
\sigma_R = 10^{-6})$, model $B$ is located at the maximum efficiency
$(\sigma_L = 0.8,\ \sigma_R = 0.2)$, and model $C$ corresponds to the
case when the FS is absent $(\sigma_L = 1,\ \sigma_R = 10^2)$.  All
three models have a substantial dynamic efficiency, but there is a
qualitative difference among them. In model $A$, ISs dissipate kinetic
to thermal energy only (thermal efficiency). In model $B$ shocks
mainly compress the magnetic field (magnetic efficiency) and dissipate
only a minor fraction of the initial kinetic and magnetic energy to
thermal energy. Finally, in the model $C$ only the RS is active.

Values $\Delta g < 2$ between adjacent parts of the flow are motivated
by the results of numerical simulations of relativistic outflows
\citep[e.g.,
][]{Aloy:2000ad,Aloy:2005zp,Mizuta:2006p1126,Zhang:2003p1193,Zhang:2004p1195,
  Morsony:2007p1208,Lazzati:2009p1210,Mizuta:2009p1213} (but see,
e.g., \cite{Kino:2008p1217}, who find $\Delta g\sim 1-19$ appropriate
to model Mrk 421). This adjacent flow regions can be approximated as
pairs of shells whose binary collision we are considering
here. However, it has been confirmed by several independent works
(e.g.,
\cite{Kobayashi:1997p657,Beloborodov:2000p632,Kobayashi:2000p599,Kino:2004p811})
that, in order to achieve a high efficiency (more than a few percent)
in internal collisions of unmagnetized shells, the ratio between the
maximum ($\Gamma_{\rm max}$) and the minimum ($\Gamma_{\rm min}$)
Lorentz factor of the distribution of initial shells should be
$\Gamma_{\rm max}/\Gamma_{\rm min}>10$.

In view of these results, we have also made an extensive analysis of
the dependence of the dynamic efficiency on $\Delta
g$. Since we are also interested in evaluating the influence of the
magnetic fields on the results, we define a new variable $\Delta s=
(1+\sigma_L)/(1+\sigma_R)$, and plot (Fig.~\ref{fig:blazar}b) the
value of the maximum efficiency reached for every combination $(\Delta
g, \Delta s)$ and fixed values of the rest of the parameters
($\Gamma_R=100$, $\chi=10^{-4}$).  We find that the maximum total
dynamic efficiency grows (non-monotonically) with increasing $\Delta
g$ (Fig.~\ref{fig:blazar}b), in agreement with the above mentioned
works for unmagnetized shell collisions. Indeed, a large value $\Delta
g \gtrsim 10$ yields dynamic efficiency values $\sim 40\%$ if both
shells are moderately magnetized ($\sigma_R\sim\sigma_L \lesssim
0.1$). Nevertheless, the amount of increase of efficiency with $\Delta
g$ depends strongly on $\Delta s$. For $|\Delta s| \gtrsim 1$,
corresponding to cases where the slower shell is highly magnetized
($\sigma_R\gtrsim 4$), the maximum dynamic efficiency is almost
independent of $\Delta g$; while for $|\Delta s| \lesssim 1$, the
maximum dynamic efficiency displays a strong, non-monotonic dependence
on $\Delta g$.

It is remarkable that values $5 \lesssim \Delta s \lesssim 100$ yield
dynamic efficiencies in excess of $\sim 20\%$, regardless of the
relative Lorentz factor between the two shells. In this region of the
parameter space the maximal dynamic efficiency happens when
$\sigma_R>10$, $\sigma_L>50$, and the total dynamic magnetic
efficiency dominates the total dynamic efficiency.

\section{Discussion}
\label{sec:discussion}

In this work we estimate the dynamic efficiency of conversion of
kinetic-to-thermal/magnetic energy in collisions (ISs) of magnetized
shells in relativistic outflows.  A fundamental difference between the
internal collisions in magnetized and unmagnetized outflows is the
fact that in the former case not only shocks but also rarefactions can
form. Thus, one would naturally expect a reduced dynamic efficiency in
the magnetized case. However, we find that the dynamic efficiency
reaches values $\sim 10\%-40\%$, in a wide range of the parameter
space typical for relativistic outflows of astrophysical interest
(blazars and GRBs). Thus, the dynamic efficiency of moderately
magnetized shell interactions is larger than in the corresponding
unmagnetized case. This is because when the shells are moderately
magnetized, most of the initial shell kinetic energy is converted to
magnetic energy, rather than to thermal energy.

The difference in efficiency between flows in blazars ($\Gamma_R=10$)
and GRBs ($\Gamma_R=100$) is very small if $\Delta g$ is fixed. From
theoretical and numerical grounds, values of $\Delta g\simeq 1$ seem
reasonable, and $\Delta g =1$ has been taken as a typical value for
both blazars and GRB jets, which brings maximum efficiency when the
magnetizations of the colliding shells are
$(\sigma_L,\sigma_R)\simeq(1,0.2)$. Larger dynamic efficiency values
$\sim 40\%$ are reached if $\Delta g\gtrsim 10$ and $|\Delta
s|\lesssim 0.1$, corresponding to cases where the magnetization of
both shells is moderate ($\sigma_R\simeq \sigma_L\lesssim 0.1$).

In the limit of low magnetization of both shells, the kinetic energy
is mostly converted into thermal energy, where the increased magnetic
energy in the shocked plasma is only a minor contribution to the total
dynamic efficiency, i.e., $\varepsilon_T \ll \varepsilon_M$. Here we
find that as the magnetization of the shells grows, the roles of
$\varepsilon_T$ and $\varepsilon_M$ are exchanged, so that
$\varepsilon_T < \varepsilon_M$ (at the maximum dynamic efficiency
$\varepsilon_T \simeq 0.1 \varepsilon_M$). If the magnetization of
both shells is large, the dynamic efficiency decreases again because
producing shocks in highly magnetized media is very difficult. All
these conclusions are independent on the EoS used to model the plasma
\cite{Mimica:2010Al}.

The comparison of our results with previous analytic or semi-analytic
works is not straightforward, since, generally, they do not compute
the (magneto-)hydrodynamic shell evolution, and they include multiple
interactions of a number of dense shells. The bottom line in these
previous works is that internal collisions of unmagnetized shells can
be extremely efficient ($\sim 100\%$; \cite{Beloborodov:2000p632}) if
the spread of the Lorentz factor (i.e., the ratio between the Lorentz
factor of the faster, $\Gamma_{\rm max}$, and of the slower
$\Gamma_{\rm min}$ shell in the sample) is large ($\Gamma_{\rm max}
/\Gamma_{\rm min}=10^3$; e.g.,
\cite{Kobayashi:1997p657,Kobayashi:2002p1216}). A more moderate spread
of the Lorentz factor $\Gamma_{\rm max}/\Gamma_{\rm min} =10$, yields
efficiencies $\sim 20\%$. These high efficiencies result after a large
number of binary collisions. Since more than a single collision is
included, the kinetic energy remaining in the flow after the first
generation of collisions, can be further converted into internal
energy as subsequent generations of collisions take place. In
contrast, we find that moderate magnetizations of both shells
($\sigma\lesssim 0.1$) and $\Delta g\gtrsim 10$ (which would roughly
correspond to $\Gamma_{\rm max}/\Gamma_{\rm min} =9$) are enough for a
single binary collision to reach a total dynamic efficiency of $\sim
40\%$.

We point out that the energy radiated in the collision of magnetized
shells is only a fraction, $f_r\lesssim 1$
\citep[e.g.,][]{Panaitescu:1999p1022,Kumar:1999p1084,Beloborodov:2000p632}
of the energy dynamically converted into thermal or magnetic
energy.Hence, a single binary collision between moderately magnetized
shells may yield radiative efficiencies $\sim 0.4f_r$. Therefore, (1)
binary collisions in relativistic magnetized flows are efficient
enough, from the dynamical point of view, to be a valid mechanism to
dissipate the bulk kinetic energy of relativistic ejecta, and (2) the
main restriction on the radiative efficiency comes from the radiation
mechanism setting the limiting factor $f_r$. The estimated dynamic
efficiency in the binary collision of magnetized shells will be
considered in a future work by computing the radiative efficiency
using the method devised in \cite{Mimica:2009p1445}.


\begin{theacknowledgments}
  MAA is an Starting Independent Grant fellow of the European Research
  Councill. We acknowledge the support from the Spanish Ministry of
  Education and Science through grants AYA2007-67626-C03-01 and
  CSD2007-00050. The authors thankfully acknowledge the computer
  resources, technical expertise and assistance provided by the
  Barcelona Supercomputing Center - Centro Nacional de
  Supercomputaci\'on.
\end{theacknowledgments}

\bibliographystyle{aipproc}   

\bibliography{nashville}

\end{document}